\documentclass[%
 reprint,
 amsmath,amssymb,
 aps,
]{revtex4-1}

\usepackage{graphicx}% Include figure files
\usepackage{dcolumn}% Align table columns on decimal point
\usepackage{bm}% bold math
\usepackage{esint}% int symbols
\begin{document}

\preprint{APS/123-QED}

\title{Nonperiodic One-gap Potentials in Quantum Mechanics}% Force line breaks with \\

\author{Dmitry V. Zakharov}
 \affiliation{Courant Institute of Mathematical Sciences, New York University\\
 251 Mercer Street, New York, NY, 10012}
%\affiliation{%
%\collaboration{MUSO Collaboration}%

\author{Vladimir E. Zakharov}
\author{Sergey A. Dyachenko}
\affiliation{
  Department of Mathematics, University of Arizona\\
  Tucson, AZ, 85791, USA
}%

\date{\today}% It is always \today, today,

\begin{abstract}
We describe a broad class of bounded non-periodic potentials in one-dimensional stationary quantum mechanics having the same spectral properties as periodic potentials. The spectrum of the corresponding Schr\"odinger operator consists of a finite or infinite number of allowed bands separated by gaps. In this letter we consider the simplest class of potentials, whose spectra consist of an interval on the negative semiaxis and the entire positive axis. The potentials are reflectionless, and a particle with positive energy moves freely in both directions. The potential is constructed as a limit of Bargmann potentials and is determined by a Riemann-Hilbert problem, which is equivalent to a pair of singular integral equations that can be efficiently solved using numerical techniques.
\begin{description}
\item[PACS numbers]{02.30.Ik, 03.65.Nk}
\end{description}
\end{abstract}

\pacs{Valid PACS appear here}% PACS, the Physics and Astronomy
\maketitle

%\tableofcontents

{\it Introduction}. We consider the Schr\"odinger equation on the real line:
\begin{equation}
-\psi'' + u(x)\psi = E\psi \quad \mbox{for $-\infty < x < \infty$}, \label{Schrodinger1}
\end{equation}
where $u(x)$ is a bounded potential. The spectrum of $u(x)$ is the set of values of $E$ for which there exists a bounded solution $\psi(x,E)$ of (\ref{Schrodinger1}).

Traditionally, the goal of spectral theory in quantum mechanics has been to describe the possible spectra of a given class of potentials of (\ref{Schrodinger1}). For an arbitrary bounded potential, the answer is not known in general. Two cases have been extensively studied and are well understood: the rapidly decaying case and the periodic case. The spectrum of a rapidly decaying potential consists of the positive real axis and a finite discrete set on the negative real axis. For a periodic potential, the spectrum is purely continuous and consists of an infinite collection of allowed bands separated by spectral gaps. Potentials having only finitely many spectral gaps form a dense subset among all periodic potentials. These potentials are described by algebro-geometric methods and are expressed by the Matveev--Its formula in terms of the Riemann $\theta$-functions of hyperelliptic curves. The Matveev--Its formula also describes a class of quasi-periodic potentials having a finite-gap structure. However, the spectrum of a generic quasi-periodic potential has a complicated Cantor set-like structure, and the corresponding system displays Anderson localization. 

In this letter, we formulate the opposite problem of describing the set of potentials of the one-dimensional Schr\"odinger operator that have a given spectrum. Specifically, we pose the question of describing the class of bounded potentials of (\ref{Schrodinger1}) having the same spectrum as a finite-gap algebro-geometric potential, but which are neither periodic nor quasi-periodic. Physically such a potential describes a non-periodic medium which nevertheless admits free wave propagation. Solving this problem is especially relevant given the recent explosion of research into metamaterias that have predetermined optical properties or a given electronic band structure.

In this letter we focus our study on one-gap potentials, with a spectrum consisting of the positive real axis and one allowed band $[-k_2^2,-k_1^2]$ on the negative real axis, where $0<k_1<k_2$. A periodic one-gap potential with such a spectrum is determined by the formula
\begin{equation}
u(x) = 2\wp(x+i\omega'-x_0)+e_3,\label{transl} \tag{1a}
\end{equation}
where $\wp(z)$ is the elliptic Weierstrass function with periods $2\omega$ and $2i\omega'$, and 
\begin{equation}
k_2^2=e_1-e_3 ,\quad k_1^2=e_2-e_3,\quad e_1+e_2+e_3=0, \label{intervals}
\end{equation}
where $e_1>e_2>e_3$ are the values of $\wp(z)$ at the half-periods of the lattice. The spectrum of the corresponding operator is doubly degenerate and reflectionless. We show that a generic one-gap reflectionless potential is determined by two positive continuous functions 
defined on the allowed band.

{\it Bargmann potentials via dressing method.} We describe an efficient method for constructing non-periodic one-gap potentials by taking the closure of the set of reflectionless  Bargmann potentials~\cite{Bargmann1949}. This construction was formally carried out  in the works of Marchenko et al~\cite{Marchenko1988}, \cite{Lundina1990} and~\cite{Marchenko1991}, but their results are not effective. We support our results with numerical simulations.

Since the discovery of the IST in 1967~\cite{Gardner1967}, the Schr\"odinger operator has been used for exact integration of the Kortweg-de Vries (KdV) equation. The reflectionless Bargmann potentials were reinterpreted as the time frames of the fundamental $N$-soliton solutions of KdV. For this reason, and also because the dressing method is local in time, we retain this terminology.

The most direct and elegant method for constructing Bargmann potentials, known as the dressing method, was described in~\cite{Zakharov1985}. The dressing method is a considerably more flexible tool than the traditional Inverse Scattering Method (ISM).

Following the cited paper, we consider the following $\bar \partial$-problem on the complex $k$-plane:
\begin{equation}
\dfrac{\partial \chi}{\partial \bar k} = ie^{2ikx}T(k)\chi(x,-k), \label{dbar}
\end{equation}
where $T(k)$ is a compactly supported distribution called the {\it dressing function} of the $\bar \partial$-problem. The $\bar \partial$-problem is normalized by the condition $\chi\to1$ as $|k|\to\infty$. The solution satisfies the following integral equation:
\begin{equation}
\chi(x,k) = 1 + \dfrac{i}{\pi}\iint\dfrac{T(-q)\chi(x,q)e^{-2iqx}}{k+q}\,dqd\bar q \label{inteqn}
\end{equation}
where we regularize the integral in the following way:
\begin{equation*}
\dfrac{1}{k}=\lim_{\varepsilon\to0}\dfrac{\bar k}{k^2 + \varepsilon^2} \quad 
\dfrac{\partial}{\partial \bar k} \dfrac{1}{k} = \pi \delta(k),
\end{equation*}
where $\delta(k)$ is the two-dimensional $\delta$-function. Suppose that the dressing function $T(k)$ has the property that equation~\eqref{inteqn} has a unique solution. Then
\begin{equation}
\chi(x,k) \to 1 + \dfrac{i\chi_0(x)}{k} + \ldots \quad\mbox{ as $|k|\to\infty$} \label{asychi}
\end{equation}
and the function $\chi$ is a solution of the equation:
\begin{equation}
\chi_{xx} - 2ik\chi_x - u(x)\chi = 0, \quad u(x) = 2\dfrac{d}{dx}\chi_0(x).\label{odeChi}
\end{equation}
To construct  Bargmann potentials, we consider the dressing function
\begin{equation}
T(k) = \pi \sum_nT_n\delta(k-i\kappa_n), \label{finiteT}
\end{equation}
where $T_n$ and $\kappa_n$ are nonzero real numbers satisfying $|\kappa_n|\in[k_1, k_2]$ and $T_n/\kappa_n > 0$ for all $n$. Then $\chi$ is a rational function:
\begin{equation}
\chi(x,k) = 1 + i\sum\dfrac{\chi_n(x)}{k-i\kappa_n}\quad\mbox{where $\chi_n(x)$ are real.} \label{chisol}
\end{equation}
The corresponding potential
\begin{equation}
u = 2\dfrac{d}{dx} \sum_{n=1}^{N} \chi_n(x)
\end{equation}
is a reflectionless Bargmann potential. The function $\psi = \chi e^{-ik x}$ satisfies the Schr\"odinger equation~\eqref{Schrodinger1}. The corresponding potential is rapidly vanishing and has a finite discrete spectrum $\{-\kappa_1^2,\ldots,-\kappa_N^2\}$, and $\psi_n(x)=\chi_n(x)e^{\kappa_n x}$ are the corresponding
eigenfunctions.

The basic relation~\eqref{dbar} is equivalent to the following linear system on the eigenfunctions:
%the equality
%\begin{equation}
%\chi_n(x)=e^{-2\kappa_n x}T_n\chi(x,-i\kappa_n), \label{eqeq1}
%\end{equation}
%which is equivalent to
\begin{equation}
\psi_n(x) + T_n\sum_{n=1}^N \dfrac{e^{-(\kappa_n+\kappa_m)x}}{\kappa_n+\kappa_m}\psi_m = T_ne^{-\kappa_n x} \label{syseq}
\end{equation}
if $T_n/\kappa_n>0$ the determinant of this system is positive
\begin{equation}
A = \det\left\| \delta_{nm} + \dfrac{T_ne^{-(\kappa_n+\kappa_m)x}}{\kappa_n+\kappa_m}\right\| > 0,
\end{equation}
so system~\eqref{syseq} has a unique solution, and moreover
\begin{equation}
u(x) = -\dfrac{d^2}{d x^2} \ln{A} \label{UviaA}
\end{equation}

{\it Closure of Bargmann potentials.} To describe the closure of the set of $N$-soliton solutions, we replace the finite dressing function (\ref{finiteT}) with one supported on two cuts $[ik_1,ik_2]$ and $[-ik_2,-ik_1]$ on the imaginary axis:
\begin{equation}
T(k) = \pi \int_{k_1}^{k_2} R_1(\kappa)\delta(k - i\kappa)\, d\kappa + \pi \int_{k_1}^{k_2}R_2(\kappa)
\delta(k+i\kappa)\,d\kappa.\label{dress_rep}
\end{equation}
Here $R_1$ and $R_2$ are functions on $[k_1,k_2]$. Formula~\eqref{chisol} transforms  into the following spectral representation:
\begin{equation}
\chi(x,k) = 1 + i\int_{k_1}^{k_2} \dfrac{f(x,p)}{k-ip}\,dp + i \int_{k_1}^{k_2}\dfrac{g(x,p)}{k+ip}\,dp \label{intrep}
\end{equation}
where $f(x,p)$ and $g(x,p)$ are real-valued functions. Then 
\begin{equation}
\chi_0 = \int_{k_1}^{k_2}\left[ f(x,p) + g(x,p)\right]dp\quad \mbox{and}\quad u(x) = 2\dfrac{d\chi_0}{dx} \label{Udef}
\end{equation}
The function $\chi(x,k)$ is analytic away from the cuts. The details of the transition from 
the discrete to continuous case are described in~\cite{ZakharovArxiv}. The algebraic system~\eqref{syseq} transforms into a system of two integral equations:
%\lipsum[1]
\begin{widetext}
\begin{align}
&f(x,k) + R_1(k)e^{-2k x}\left[\int_{k_1}^{k_2} \dfrac{f(x,q)}{k+q}\,dq + 
 \fint_{k_1}^{k_2} \dfrac{g(x,q)}{k-q}\,dq\right] = R_1(k)e^{-2k x} \label{eq1s} \\
&g(x,k) + R_2(k)e^{2k x}\left[\fint_{k_1}^{k_2} \dfrac{f(x,q)}{k-q}\,dq + 
 \int_{k_1}^{k_2} \dfrac{g(x,q)}{k+q}\,dq\right] = -R_2(k)e^{2k x} \label{eq2s} 
\end{align}
\end{widetext}
%\lipsum[1]
In~\cite{ZakharovArxiv} we show that if $R_1$ and $R_2$ are continuous, positive and $\alpha$-H\"older for a positive $\alpha$, then system~\eqref{eq1s}-\eqref{eq2s} has a unique solution. Moreover,
\begin{equation}
\varphi(x,k) = f(x,k)e^{ik x}\quad\mbox{and}\quad \psi(x,k)=g(x,k)e^{-ik x} \label{eqnfg}
\end{equation}
are eigenfunctions of the Schr\"odinger operator~\eqref{Schrodinger1} corresponding to $E = k^2$. These eigenfunctions form an orthogonal collection:
\begin{align}
&\int_{-\infty}^{\infty} \varphi(x,\kappa)\varphi(x,\kappa')\,d\kappa = R_1\delta(\kappa-\kappa'), \label{orth1} \\
&\int_{-\infty}^{\infty} \psi(x,\kappa)\psi(x,\kappa')\,d\kappa = R_2\delta(\kappa-\kappa'), \label{orth2} \\
&\int_{-\infty}^{\infty} \psi(x,\kappa)\varphi(x,\kappa')\, d\kappa = 0.  \label{orth3}
\end{align}

On any interval where one of the functions $R_1$, $R_2$ is positive and the other vanishes,  the spectrum is simple. If $R_1$ and $R_2$ both vanish, the corresponding segment is a spectral gap, so the finite-gap case is included in our description. If $R_1= R_2$, then $u(x) = u(-x)$ and the potential is symmetric. 

We also show in~\cite{ZakharovArxiv} that system~\eqref{eq1s}-\eqref{eq2s} is equivalent to the following Riemann--Hilbert problem on $\chi$:
\begin{align}
\begin{split}
&\chi^{+}(x,ik) - \chi^{-}(x,ik) = \\
&\qquad =i\pi R_1(k)e^{-2k x}\left[\chi^{+}(x,-ik) + \chi^{-}(x,-ik) \right], \label{relatcuts1}
\end{split}
\end{align}
\begin{align}
\begin{split}
&\chi^{+}(x,-ik) - \chi^{-}(x,-ik) = \\
&\qquad -i\pi R_2(k)e^{2k x}\left[\chi^{+}(x,ik) + \chi^{-}(x,ik) \right], \label{relatcuts2}
\end{split}
\end{align}
where $\chi^{\pm}$ are the boundary values of $\chi$ along the cuts:
$$
\chi^{\pm}(x,k)=\displaystyle\lim_{\varepsilon\to 0} \chi(x,k\pm i\varepsilon),\quad k\in [-ik_2,-ik_1]\cup
[ik_1,ik_2].
$$

A fundamental difference between the ISM and the dressing method is that in the former, the scattering data can be uniquely restored from the potential, while in the latter, the same potential can be constructed using a variety of different dressings. For example, in t (\ref{finiteT}) we can change the signs of one of the $T_n$, and the corresponding $\kappa_n$, without changing $u(x)$ (see~\cite{ZakharovArxiv} for details), so each $N$-soliton solution can be constructed using $2^N$ different dressing functions of the form (\ref{finiteT}) . Hence we cannot expect a given bounded potential to be determined by a unique choice of $R_1$ and $R_2$. Below we describe a class of dressings leading to one-gap periodic potentials.

\begin{figure*}[ht]
\includegraphics[width=0.32\textwidth]{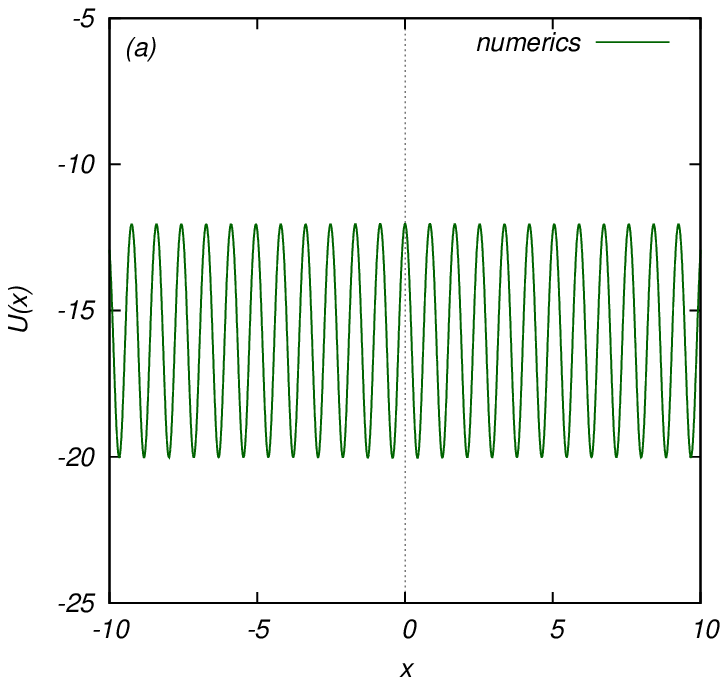}
\includegraphics[width=0.32\textwidth]{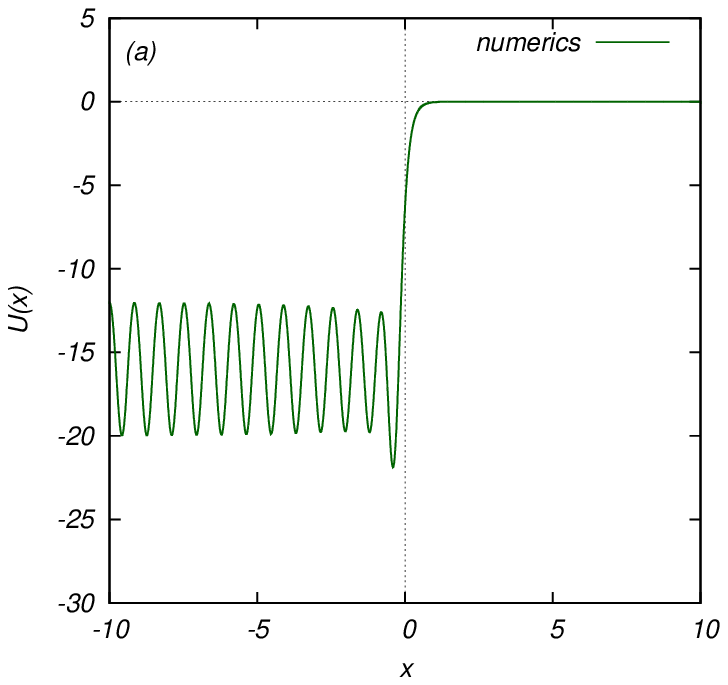}
\includegraphics[width=0.32\textwidth]{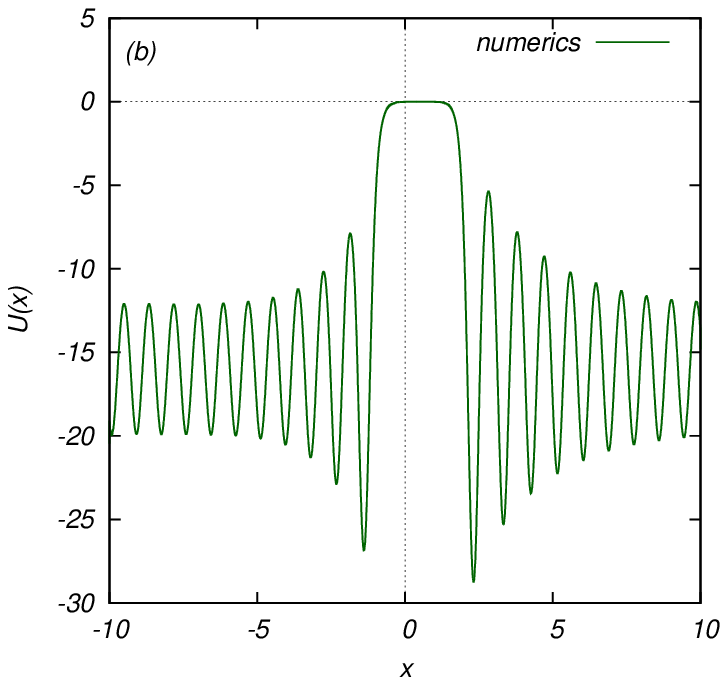}
\caption{A potential $U(x)$ that appears as a result of dressing with: (a) $R_1 = R_2 = 1/\pi$ (periodic); (b) $R_1 = 1/\pi$ and $R_2 = 0$ decaying for $x>0$; 
(c)$R_1 = \frac{1}{\pi}\times10^{-3}$ and $R_2 = \frac{1}{\pi}\times10^{-6}$. }
\label{fig3}
\end{figure*}

{\it Periodic potentials.} Suppose that a potential $u(x)$ evolves according to some equation of the KdV hierarchy (see~\cite{Novikov1984}). The dressing functions evolve as follows:
\begin{equation}
R_1(k) \to R_1(k)e^{S(k)t},\quad 
R_2(k) \to R_2(k)e^{-S(k)t},\label{evol_op} 
\end{equation}
where $S(k)$ is an odd function (for example, for KdV $S(k) = 8k^3t$). A periodic potential remains periodic at any moment of time, so applying transformation~\eqref{evol_op} does not change the periodicity properties of the potential.  We illustrate this with a numerical simulation.

We now construct a dressing function of the form (\ref{dress_rep}) that determines a periodic potential. We put $x_0 = \omega$ in~\eqref{transl} and map the $k$-plane to the period parallelogram as follows:
\begin{equation}
k^2 = e_3 - \wp(z) \quad z(k)\to -\dfrac{i}{k}\quad\mbox{as}\quad |k|\to\infty. \label{pmap}
\end{equation}
Then~\eqref{Schrodinger1} becomes the Lam\'e equation
\begin{equation}
\varphi'' - \left[ 2\wp(x-\omega'-i\omega') + \wp(z)\right]\varphi = 0 \label{Lame}
\end{equation}
which has following solution:
\begin{equation}
\varphi(x,z) = \dfrac{\sigma(x-\omega-i\omega + z)\sigma(\omega + i\omega')}{\sigma(x-\omega-i\omega)\sigma(\omega + i\omega' -z)}\exp{(-\zeta(z)x)}. \label{phisol}
\end{equation}

Now define the function:
\begin{equation}
\xi(x,k) = \dfrac{k-ik_1}{k-ik_2}\varphi(x,z(k))e^{ikx}. \label{xi}
\end{equation}
This function satisfies the equation:
\begin{equation}
\xi''-2ik\xi'-u(x)\xi = 0, \quad\mbox{for}\quad \xi\to 1\quad |k|\to\infty \label{eqxi}
\end{equation}
The function $\xi$ satisfies the RH problem~\eqref{relatcuts1}-\eqref{relatcuts2}, where:
\begin{align}
&R_1(q) = \dfrac{1}{\pi}h(q), \quad R_2(q) = \dfrac{1}{\pi h(q)}, \label{RH1} \\
&h(q) = \sqrt{\dfrac{(k_2-q)(q+k_1)}{(q-k_1)(q+k_2)}}, \quad k_1< q < k_2. \label{hofq}
\end{align}
Now $S(q)=\ln{h(q)}$ is an odd function, so we can replace $h(q)$ with $1$ and 
assume that
\begin{equation}
R_1(q) = R_2(q) = \dfrac{1}{\pi}. \label{periodicdress}
\end{equation}
The dressings~\eqref{RH1}-\eqref{hofq} and~\eqref{periodicdress} both define a periodic potential.

%\section{\label{sec:level5}Numerical Simulation.}
{\it Numerical solution.} We solve equations~\eqref{eq1s}-\eqref{eq2s} numerically for $k_1 = 2$ and $k_2 = 4$. It is convenient to replace $\phi(x,k)$ and $\psi(x,k)$ with the following functions:
\begin{align*}
P(x,k) = \sqrt{1-k^2}\phi(x,k), \\
Q(x,k) = \sqrt{1-k^2}\psi(x,k), 
\end{align*}
which then satisfy the following integral equations: 
\begin{widetext}
\begin{align}
&P(x,k)+ r_1(x,k)\left[\int_{k_1}^{k_2} \dfrac{P(x,q)e^{-qx}\,dq}{(k+q)\sqrt{1-q^2}} + 
 \fint_{k_1}^{k_2} \dfrac{Q(x,q)e^{qx}\,dq}{(k-q)\sqrt{1-q^2}}\right] = r_1(x,k), \label{eq1n} \\
&Q(x,k) + r_2(x,k)\left[\fint_{k_1}^{k_2} \dfrac{P(x,q)e^{-qx}\,dq}{(k-q)\sqrt{1-q^2}} + 
 \int_{k_1}^{k_2} \dfrac{Q(x,q)e^{qx}\,dq}{(k+q)\sqrt{1-q^2}}\right] = -r_2(x,k), \label{eq2n} 
\end{align}
\end{widetext}
where $r_j(x,k) = \sqrt{1-k^2}R_j(k)e^{(-1)^jkx}$ for $j = 1,2$.
Discretized at Chebyshev nodes $q_k =  \cos{\frac{ (2k-1)\pi}{2M}}$ with $k = 1,2,\ldots,M$ the integrals are evaluated via
Gauss-Chebyshev quadrature that is exact for polynomials of degree less than $2M-1$. Note that each equation of the
system contains a Cauchy principal value integral denoted by $\fint$, and that integration in the vicinity the of singularity at $q = k$ requires a
shift from real axis. %For numerical purpose a small but finite shift is done instead.

The spatial variable $x$ appears as a parameter in~\eqref{eq1n}-\eqref{eq2n} and the $x$-dependence of $r_1$ and $r_2$ becomes a major obstacle, 
since the condition number of the discretized system is exponential in $x$ and requires usage of multiprecision arithmetics. 
%We use a brute-force approach and increase the number of digits via using multiprecision capabilities of the Julia computing language.

A mesh for the spatial parameter $x$ is a Chebyshev grid with $M$ nodes; a high-order polynomial interpolation by means of
Lagrange interpolation is used for intermediate points. In a typical simulation an interpolating polynomial of degree $200$ suffices to have an accurate approximation 
for $|x|< 10$ (See figures~\ref{fig3}).

{\it Conclusion.} The procedure of closing the set of Bargmann potentials described above can be generalized to a wide class of linear operators, such as the Dirac operator. This is especially important for various nonlinear integrable systems, such as the Nonlinear Sch\"odinger Equation and the Kadomtsev--Petviashvili equation, where these linear operators play a fundamental role and which are solved using the ISM. 

To apply the theory that has been developed for these systems to real physical problems, it is necessary to develop a statistical theory of integrable systems with an infinite number of degrees of freedom. This theory of integrable turbulence is still very much in its infancy, the first stages have been suggested in~\cite{Zakharov2009}. The technique described in this letter shows an approach to constructing strongly nonlinear statistically homogeneous solutions to integrable systems such as KdV and the nonlinear Schr\"odinger equation. In fact, integrable turbulence is a common physical phenomenon. This turbulence takes place in the coastal areas of seas, and describes effects occurring in optical fibers. Thus the natural extension and our next step will be the study of random non-periodic potentials with a continuous spectrum, with this work being at the core of the theory of integrable turbulence.

{\it Acknowledgments.} The second author gratefully acknowledges the support of the National Science Foundation (Grant No.1130450).


\begin{thebibliography}{99}

\bibitem{Bargmann1949}

V. Bargmann, Rev. Mod. Phys, {\bf 21}, 494, 1949

\bibitem{Marchenko1988}

V. Marchenko, {\it Nonlinear equations and operator algebras,} Mathematics and its Applications (Soviet Series), {\bf 17}, 1988

\bibitem{Lundina1990}

D. Lundina, J. Soviet Math., {\bf 3}, 290, 1990

\bibitem{Marchenko1991}

V. Marchenko, {\it What is integrability?,} 273, Springer Ser. Nonlinear Dynam., 1991


\bibitem{Gardner1967}

C.S. Gardner, J.M. Greene, M.D. Kruskal and R.M. Miura, Phys. Rev. Lett., {\bf 19}, 1095, 1967

\bibitem{Zakharov1985}

V. Zakharov and S. Manakov, Funktsional. Anal. i Prilozhen. (in Russian), {\bf 19}, 11, 1985

\bibitem{ZakharovArxiv}

D. Zakharov and V. Zakharov, to appear on arxiv, 2015

\bibitem{Novikov1984}
S. Novikov, S. Manakov, L. Pitaevskii and V. Zakharov,
{\it Theory of solitons. The inverse scattering method.}, Contemporary Soviet Mathematics, 1984

\bibitem{Zakharov2009}

V. Zakharov, Stud. Appl. Math., {\bf 122}, 3, 219, 2009

\end{thebibliography}
\end{document}